\documentclass[twocolumn,prb,amsmath,amssymb,showpacs]{revtex4}

\usepackage{amsmath}
\usepackage{amsxtra,amscd,upref,layout,bm}
\usepackage{calc}
\usepackage{graphicx,color}

\newcommand{\eref}[1]{Eq.~(\ref{#1})}
\newcommand{\eeref}[2]{Eqs.~(\ref{#1},\ref{#2})}
\newcommand{\eeeref}[3]{Eqs.~(\ref{#1},\ref{#2},\ref{#3})}

\newcommand{\sref}[1]{Sec.~\ref{#1}}
\newcommand{\cref}[1]{Chap.~\ref{#1}}
\newcommand{\tref}[1]{Tab.~\ref{#1}}
\newcommand{\fref}[1]{Fig.~\ref{#1}}

\newcommand{\degC}{{\ensuremath{{}^{\mathrm{o}}}}}

\newcommand{\Hvh}{{\ensuremath{{\widehat{\bm{h}}}}}}
\newcommand{\VK}[1]{{\ensuremath{{\bm{#1}}}}}

\newcommand{\Qv}{{\ensuremath{{\bm{q}}}}}
\newcommand{\Hv}{{\ensuremath{{\bm{h}}}}}
\newcommand{\Kv}{{\ensuremath{{\bm{k}}}}}
\newcommand{\Rv}{{\ensuremath{{\bm{r}}}}}

\newcommand{\Vv}{{\ensuremath{{\bm{v}}}}}

\begin{document}                  

     
     

\title{Nanoparticle size distribution estimation by full-pattern powder 
diffraction analysis.}

\author{A. Cervellino} 
 \altaffiliation[On leave from]{%
 Consiglio Nazionale delle Ricerche, Istituto 
 di Cristallografia (CNR-IC),
 Via Amendola 122/O, I-70126 Bari, Italy.
 }
\affiliation{%
Laboratory for Neutron Scattering, Paul Scherrer Institute,  
CH-5232 Villigen PSI, Switzerland
}
\author{C. Giannini}
 \email{cinzia.giannini@ic.cnr.it}
\author{A. Guagliardi and M. Ladisa}
\affiliation{
Consiglio Nazionale delle Ricerche, Istituto 
 di Cristallografia (CNR-IC), 
 Via Amendola 122/O, I-70126 Bari, Italy.
}%

\date{\today}

\begin{abstract}
    The increasing scientific and technological interest in 
nanoparticles has 
    raised the need for fast, efficient and precise characterization 
techniques. 
    Powder diffraction is a very efficient experimental method, as it 
    is straightforward and non-destructive. However, its use for 
    extracting 
    information regarding very small particles brings some common 
    crystallographic approximations to and beyond their limits 
    of validity. 
    Powder pattern diffraction calculation methods are 
    critically discussed, with special focus on 
    spherical particles with log-normal 
    distribution, with the target of determining size distribution parameters. 
    A 20-nm CeO$_{2}$ sample is analyzed as example. 
\end{abstract}

\maketitle

\section{Introduction}\label{S:intro}

We are assisting at a booming expansion of nanoparticle research and 
technology. 
Synthesis method especially make fast progresses\cite{Masala04}. 
Analysis methods, however, are not up to speed. A fundamental 
simple task as determining and controlling the size distribution of 
nanoparticles (NPs hereafter) is currently a complex experimental work, 
involving electron microscopy and combined techniques. In this work we 
want to highlight the possibilities offered in this issue by a much 
less complex technique as powder diffraction.

Powder diffraction is a widespread technique with a great potential 
to meet 
the increasing demands of microstructural material 
characterization. The methods of powder diffraction data analysis 
have 
reached maturity for micrometer-sized polycrystalline 
materials. However, when the particle size falls much below 
100 nm, specifically tuned methods of analysis are needed to extract 
meaningful information from powder diffraction patterns. 
In fact, nanoparticles (NPs hereafter) present 
unique analytical challenges. 
In the most complex cases, non-crystallographic structures 
\cite{Ino66,Ino67,Ino69a,Ino69,Marks94,Vogel98,Zanchet2000,Hall2000,NOI03,
NOI04} may 
occur. 
Surface-related deformation fields 
\cite{Palosz2001,Palosz2003,Palosz2004}
are another challenge. In these extreme cases, the classical 
crystallographic
formalism becomes quite useless. 
The Debye scattering function\cite{GuinierDEB} 
(that is, the direct evaluation of the NP structure factor from the 
interatomic distances) 
is the only choice in those cases. We are currently developing  
\cite{NOI03,NOI05}  
methods to increase the efficiency of such calculations and make them 
a
practical tool.

Even for crystalline NPs, however, the small size plays a decisive 
role. Bragg peaks may be so much broadened that they cannot be simply 
separated and many approximations, commonly accepted for micrometer 
size 
domains, fail. 
As we will show, also models specifically 
corrected for NPs\cite{InoMinami79,MinamiIno79,InoMinami84} may fail 
for ultra-small NPs (say below 5 nm diameter, as it will be better 
specified). Again for these ultra-small sizes the Debye scattering 
function is the only choice for obtaining precise results, while 
the smaller number of atoms makes it extremely practical.

The plan of the paper is the following. 
In \sref{Sec1} we discuss the shape-based method for calculating 
NP powder patterns in relation to the surface structure and 
to its limits of validity at small sizes. 
Application to full-pattern fit on a 
test-case (20-nm CeO${}_2$) is shown in \sref{Sec3}.
\footnote{Other NP samples (ZnTe, ZnSe) in the 5--20~nm size range 
have been treated successfully and the results shall 
be published separately.}
Summary and conclusions are given in \sref{Sec4}.

\section{Powder patterns and size information}\label{Sec1}

Scherrer's formula\cite{Scherrer18} is the most known method 
for extracting size information from powder patterns 
(namely, from the Bragg peaks' width). 
This is a simple method, but accurate only to the order of magnitude. 
However, 
since Scherrer's work, line profile analysis has made 
enormous 
progress
\cite{WarrenAverbach52,Warren69,Wilkens70,Langford78,%
Ungar98,Langford00,Ungar01,Groma01,Marinkovic2001,Ungar02}. 

Theoretical progress on understanding the physical origin of peak 
broadening has been focused 
on the dislocation analysis, size broadening being considered 
as a side effect to be corrected for in order to determine the 
defect structure. Nevertheless, 
today it is possible to determine the parameters of a (log-normal) 
size distribution of crystallites, together with information on type 
and concentration of dislocations. These methods are, however, 
complex 
and sophisticated, requiring a fairly high signal-to-noise ratio, low 
and flat background, a precise deconvolution of the instrumental 
broadening and especially well-isolated Bragg peaks. 

Full-pattern fitting methods (\emph{cf.} \sref{Sec2}) are more 
direct and robust, especially when the target is the size analysis. 
Firstly, they use all the experimental information, 
regardless of partial or total peak overlap, 
increasing redundancy and therefore precision and decreasing 
experimental requirement. 
Furthermore, they allow 
the evaluation of a NP-characteristic feature, namely the 
variation with size of the lattice parameter\cite{NOI03,NOI04} (an 
effect 
that can be important below 20~nm). 
Corrections for texture, microabsorption, anisotropic elastic peak 
shifts 
and instrumental broadening 
can also be implemented. 

An efficient and precise method to evaluate NP diffraction patterns 
is needed to perform full-pattern fits. 
Hereafter we discuss the shape-based 
method\cite{InoMinami79,MinamiIno79,InoMinami84} 
with a thorough analysis of its validity limits.

\subsection{NP shape-based diffraction models}\label{Sec2}

We shortly recall some methods for the calculation of the powder 
diffraction intensity for a NP with known periodic structure and 
definite size and shape. 
In the following the length of a vector $\Vv$ will be 
denoted by $v$. Accordingly, $\Qv$ will be the scattering vector of 
length 
$q=2\sin\theta/\lambda$, where $\theta$ is the scattering half-angle 
and $\lambda$ the incident wavelength; $\Hv$ shall denote the 
scattering vector  
associated with a Bragg peak, its length being $h$. 
A NP occupies a geometrical region of space $G$. 
We recall \cite{Patt39s,Ewald40} the definition of a shape 
function $S(\Rv)$, such that $S(\Rv)=1$ if $\Rv$ lies inside $G$, 
$S(\Rv)=0$ otherwise. 
We shall hereforth suppose that $S(-\Rv)=S(\Rv)$ so that its Fourier 
transform is real. 

However, defining the shape of a 
crystal means also to describe what happens to the atoms on the 
surface. 
These are increasingly important at very small sizes. 
In fact, there are different ways of interpreting the action of 
$S(\Rv)$, the most meaningful ones being: 
\begin{itemize}
\item[a)\ ]{truncating sharply the scattering density (the electron 
density for 
x-rays) at the surface \cite{Patt39s,Ewald40};}
\item[b)\ ]{selecting all whole unit cells whose origins are in $G$ 
and all 
whole atoms whose centres lie in the selected cells\cite{HoseBag62};}
\item[c)\ ]{selecting all whole atoms whose centres are in $G$.}
\end{itemize}
Useful illustrations are found in Fig.~1 of 
Ref.~\onlinecite{InoMinami79} (see Figs.~1a, 1c and 1d, 
respectively for a, b, c). 
\footnote{We remark here that constructions b) and c) are less 
dramatically 
different than what appears in 
\onlinecite{InoMinami79}, depending on a proper choice of the unit 
cell. 
A physically descriptive choice, as the 
Wigner-Seitz unit cell, for instance, would reduce those 
differences.}
To evaluate the diffracted intensities, in cases b), c), one may 
utilize the Debye function. In this way the chosen model is 
faithfully represented. It is possible, however, to proceed in a 
different way, that is, by the shape-function method. 
Accordingly, we first evaluate the scattering amplitude $A(\Qv)$. 
The explicit expressions\cite{InoMinami79} are, for cases a,b,c:
\begin{eqnarray}
     A^{\mathrm{a}}(\Qv) & = & 
    \mathop{\sum}_{\Hv\in\Lambda^*}\,\tilde{S}(\Qv-\Hv)F(\Hv),
    \label{eq:amplA}  \\
     A^{\mathrm{b}}(\Qv) & = & 
    \mathop{\sum}_{\Hv\in\Lambda^*}\,\tilde{S}(\Qv-\Hv)F(\Qv),
    \label{eq:amplB}  \\
     A^{\mathrm{c}}(\Qv) & = & 
    \mathop{\sum}_{\Hv\in\Lambda^*}\,\tilde{S}(\Qv-\Hv)F(\Hv,q),
    \label{eq:amplC}  
\end{eqnarray}
where $\Lambda^*$ is the reciprocal lattice; 
$\tilde{S}(\Qv)$ is the Fourier transform\footnote{For the sake of 
brevity we omit the factors $v_{c}^{-1}$ from the Fourier amplitudes 
and $v_{c}^{-2}$ from the related intensities, where $v_{c}$ is the 
unit cell volume.}
of $S(\Rv)$, or
\begin{equation}
    \tilde{S}(\Qv)=\int_{\mathbb{R}^3}\mathrm{d}^3\Rv
    S(\Rv)\mathrm{e}^{2\pi i \Qv\cdot\Rv}=\int_{G}\mathrm{d}^3\Rv \, 
\mathrm{e}^{2\pi i \Qv\cdot\Rv}\ ,
    \label{eq:FTdef}
\end{equation}
and it satisfies $\tilde{S}(\Qv)=\tilde{S}(-\Qv)$ because 
$S(-\Rv)=S(\Rv)$; 
$F(\Hv)$ is the unit cell structure factor 
\begin{equation}
F(\Hv)=\mathop{\sum}_{\alpha=1}^{N_a}f_{\alpha}(h)\mathrm{e}^{2\pi 
    i \Hv\cdot\Rv_{\alpha}}\ ,
    \label{eq:STRF}
\end{equation}
where the sum index $\alpha$ runs on the atoms in the unit cell, 
which have form factors $f_{\alpha}$ and position vectors (relative 
to the cell origin) $\Rv_{\alpha}$; 
$F(\Qv)$ is the same as the former but evaluated in $\Qv$; 
and $F(\Hv,q)$ is the mixed expression
\begin{equation}
F(\Hv,q)=\mathop{\sum}_{\alpha=1}^{N_a}f_{\alpha}(q)\mathrm{e}^{2\pi 
    i \Hv\cdot\Rv_{\alpha}}\,.
    \label{eq:STRX}
\end{equation}

It is evident that form a) is simpler but by 
construction less reasonable - for electron and x-ray diffraction - 
than 
b) and c). In fact, the sharp truncation of the electron density at 
the surface is unjustified. 
For neutron nuclear elastic scattering the atoms 
are point scatterers, therefore, construction a) coincides with c). 
Accordingly, in the neutron case, 
the atomic form factors are constant and 
$A^{\mathrm{a}}(\Qv)=A^{\mathrm{c}}(\Qv)$. 

Form b) depends on an appropriate choice 
of the unit cell. Clearly, it preserves the stoichiometric 
composition and symmetry. 

Form c) needs a careful implementation (regarding the definition of 
$G$) 
to preserve stoichiometry, that 
is important for ionic compounds; however, it is clearly more 
flexible. 
Remark also that,
in the case of monoatomic lattices, instead - as for simple-cubic, 
face-centered or body-centered cubic metals - construction b) and c) 
will be coincident and $A^{\mathrm{b}}(\Qv)=A^{\mathrm{c}}(\Qv)$.

\subsection{NP scattering intensities}\label{Sec22}

Squaring \eeeref{eq:amplA}{eq:amplB}{eq:amplC} we 
obtain the intensities. Supposing $S$ centrosymmetric and $\tilde{S}$ 
real, we have 
\begin{eqnarray}
     I^{\mathrm{a}}(\Qv) & = & 
\mathop{\sum}_{\Hv\in\Lambda^*}\,\tilde{S}^2(\Qv-\Hv)\left|F(\Hv)\right|^2,
\label{eq:intA}  \\
     I^{\mathrm{b}}(\Qv) & = & \left|F(\Qv)\right|^2
    \mathop{\sum}_{\Hv\in\Lambda^*}\,\tilde{S}^2(\Qv-\Hv),
    \label{eq:intB}  \\
     I^{\mathrm{c}}(\Qv) & = & 
\mathop{\sum}_{\Hv\in\Lambda^*}\,\tilde{S}^2(\Qv-\Hv)\left|F(\Hv,q)\right|^2.
\label{eq:intC}
\end{eqnarray}
Here, we have neglected cross-summations of the form
\begin{equation}
     \mathcal{R}(\Qv) = 
\mathop{\sum}_{\genfrac{}{}{0pt}{2}{\Hv,\Kv\in\Lambda^*}{\Kv\neq\Hv}}
     \tilde{S}(\Qv-\Hv)\tilde{S}(\Qv-\Kv) 
M_{\Qv,\Hv}^{\mathrm{x}}\overline{M}_{\Qv,\Kv}^{\mathrm{x}}
    \label{eq:X}
\end{equation}
where overbar stands for complex conjugate and, for x=a,b,c, 
respectively, it is $M_{\Qv,\Hv}^{\mathrm{a}}=F(\Hv)$, 
$M_{\Qv,\Hv}^{\mathrm{b}}=F(\Qv)$ or 
$M_{\Qv,\Hv}^{\mathrm{c}}=F(\Hv,q)$. 
Neglecting $\mathcal{R}(\Qv)$ is, first of all, a question of convenience, 
because its evaluation - either analytical or numerical - is a 
nightmare. 

There are obvious reasons for neglecting $\mathcal{R}(\Qv)$ for large 
particles. 
Consider a spherical particle with cubic structure with 
lattice parameter $a$ and radius $R\gg a$. 
$\tilde{S}(\Qv)$ is large only for $q\lesssim{}1/R$, and decreases as 
$(2\pi{}qR)^{-2}$ for $q\gg 1/R$. As for any Bragg peak $\Hv$ it 
is $1/R\ll 1/a \lesssim h$, 
$\mathcal{R}(\Qv)\sim O((R/a)^{-2})$ can be neglected. 

For smaller particles the situation is different. 
In Refs.~\onlinecite{HoseBag62,InoMinami79} it is proposed that 
$\mathcal{R}(\Qv)$ is negligible due to a certain statistical `smearing' 
of the NP 
surface region on a thickness of the order of the lattice parameter 
$a$. 
However, 
this hypothesis cannot be accepted by default. 

Firstly, the order at the surface 
strongly depends on the considered 
crystal phase and on the 
actual sample. Consider that for a NP of diameter $D=Na$, 
the fraction of atoms included in a 
layer of thickness $a$ is $\approx 6/N$ 
(about 50\%{} at $D=10a$, still 12\%{} at $D=50a$). 
The structure of this large fraction should be 
carefully considered on a 
case-by-case basis. 
Relaxations in the core due to a disordered layer of thickness 
$a$ should also be considered. 
Secondly, supposing a default smearing of the NP boundaries 
flattens the different construction principles of forms a, b, c. 
In fact, the differences among them regard the finest 
details of the NP surface structure. 

We shall hereafter 
assess the effect of neglecting $\mathcal{R}(\Qv)$ on the calculation of 
a powder diffraction pattern. In App.~\ref{AppA} we carry out some 
relevant calculations. Evidently this will depend on the choice of 
form a, b, or c. Examples 
are reported in the following section. 

For form 
$I^{\mathrm{b}}(\Qv)$ it turns out that, 
even when $\mathcal{R}(\Qv)$ is not negligible, it yields a 
contribution that is approximately proportional to the retained term 
$I^{\mathrm{b}}(\Qv)$ of the scattered intensity. This means that 
the effect of neglecting $\mathcal{R}(\Qv)$ may be just a small error 
on the global scale factor for samples composed of particles of 
equal size. However, as this effect is size-dependent, it may hamper 
the 
evaluation of size distribution when this is not very narrow. 
A size-related correction factor for 
the scale factor may - and should - be evaluated (see 
App.~\ref{AppA}) in this case. This of course is an undesired 
complication. 

In cases a) and c) the neglected term $\mathcal{R}(\Qv)$ depends on 
the crystal structure (see App.~\ref{AppA}). 
It is not a constant scale factor for all Bragg peaks, and it may have 
a significant gradient in the Bragg peak positions. 
At very small sizes the latter may induce a systematic error 
also in the lattice constant determination. 
However, in the x-ray case, for form a) $\mathcal{R}(\Qv)$ is larger 
- and has a larger gradient in the Bragg peak neighbourhood -  
than the corresponding term for form c).

\subsection{NP powder patterns}\label{Sec23}

To obtain a powder diffraction pattern, we must integrate 
$I^{\mathrm{x}}(\Qv)$ (x=a,b,c, see 
\eeeref{eq:amplA}{eq:amplB}{eq:amplC}) 
at constant $q$. We write $\Qv$ in polar coordinates as 
$\Qv=(q\sin\psi\cos\phi,q\sin\psi\sin\phi,q\cos\psi)\equiv(q,\omega)$, 
where $\omega$ is the orientation 
defined by the pair $(\psi,\phi)$. 
We have to integrate over the set of all orientations 
$\Omega\equiv\{0<\psi<\pi,0<\phi<2\pi\}$ 
(with $\mathrm{d}\omega\equiv\sin\psi\mathrm{d}\psi\mathrm{d}\phi$), 
as 
\begin{equation}
\sin\psi\mathrm{d}\psi\int_{0}^{2\pi}\mathrm{d}\phi
    I^{\mathrm{x}}_{p}(q)=q^2\int_{\Omega}\mathrm{d}\omega
    I^{\mathrm{x}}(q,\omega).
    \label{eq:Ip1}
\end{equation}
In detail, considering the expressions for the different cases, we 
have
\begin{eqnarray}
    I^{\mathrm{a}}_{p}(q) & = & 
    q^2\mathop{\sum}_{\Hv\in\Lambda^*}\,
\left|F(\Hv)\right|^2\,\int_{\Omega}\mathrm{d}\omega
    \tilde{S}^2(\Qv-\Hv);
    \label{eq:intAp}  \\
        I^{\mathrm{b}}_{p}(q) & = & 
    q^2\mathop{\sum}_{\Hv\in\Lambda^*}\,\int_{\Omega}\mathrm{d}\omega
    \left|F(\Qv)\right|^2\tilde{S}^2(\Qv-\Hv);
    \label{eq:intBp}  \\
    I^{\mathrm{c}}_{p}(q) & = & 
    q^2\mathop{\sum}_{\Hv\in\Lambda^*}\,\left|F(\Hv,q)\right|^2\,
    \int_{\Omega}\mathrm{d}\omega \tilde{S}^2(\Qv-\Hv).
    \label{eq:intCp}
\end{eqnarray}
The integration in case b) is much more difficult and it cannot 
generally be expressed in closed form 
even for simple shapes. Therefore, as a careful implementation of 
form 
c) is at least as good a description as form b), 
we shall disregard b) in the following.
Suppose now that $G$ is a sphere of radius $R$ and volume $V=4\pi 
R^3/3$, we have
\begin{equation}
    \tilde{S}(\Qv)=\tilde{S}(q)=3V
\left[\frac{\sin(y)-y\cos(y)}{y^3}\right]_{y=2\pi 
q R}
    \label{eq:Ip2}
\end{equation}
and, as $|\Qv-\Hv|=(q^2+h^2-2qh\cos\psi)^{1/2}$,  
\begin{eqnarray}
    \tilde{S}(\Qv-\Hv)&=&\displaystyle{
    3V\left[\frac{\sin(y)-y\cos(y)}{y^3}\right]}
    \label{eq:Ip3}\\
    \text{with}&&{y=2\pi (q^2+h^2-2qh\cos\psi)^{1/2} R}.\nonumber
\end{eqnarray}
Substituting in \eeref{eq:intAp}{eq:intCp} yields \cite{Patt39s}
\begin{eqnarray}
    I^{\mathrm{a}}_{p}(q) 
    & = & 
    \frac{3qVR}{8\pi}\mathop{\sum}_{\Hv\in\Lambda^*}\,\left|F(\Hv)\right|^2\,
    \frac{(A_{-}-A_{+})}{h};
    \label{eq:intAp2}  \\
    I^{\mathrm{c}}_{p}(q) 
    & = & 
    \frac{3qVR}{8\pi}\mathop{\sum}_{\Hv\in\Lambda^*}\,\left|F(\Hv,q)\right|^2\,
    \frac{(A_{-}-A_{+})}{h}
    ,
    \label{eq:intCp2}\\
    \text{where} &A_{\pm}&\equiv 
    y^{-2}\left(1-\sin(2y)/y+\sin^2(y)/y^2\right)\nonumber\\
    &\text{for}&y=2\pi R(q\pm h).
    \nonumber
\end{eqnarray}
Now we consider the crystal's Laue group $\mathcal{G}$ so that we can 
extend the summation on the asymmetric part  
$\Lambda^*/\mathcal{G}$ of the reciprocal lattice:
\begin{eqnarray}
    I^{\mathrm{a}}_{p}(q) 
    & = & 
    \frac{3qVR}{8\pi}\mathop{\sum}_{\Hv\in\Lambda^*/\mathcal{G}}
\!\!\!\!\mu_{\Hv}\left|F(\Hv)\right|^2
    \frac{(A_{-}-A_{+})}{h};
    \label{eq:intAp3}  \\
    I^{\mathrm{c}}_{p}(q) 
    & = & 
    \frac{3qVR}{8\pi}\mathop{\sum}_{\Hv\in\Lambda^*/\mathcal{G}}
\!\!\!\!\mu_{\Hv}\left|F(\Hv,q)\right|^2
    \frac{(A_{-}-A_{+})}{h},
    \label{eq:intCp3}
\end{eqnarray}
where $\mu_{\Hv}$ is the multiplicity of $\Hv$ subject to 
$\mathcal{G}$. 
Evaluation of $I^{\mathrm{c}}_{p}(q)$ is only slightly more complex 
than $I^{\mathrm{a}}_{p}(q)$, 
and the gain in accuracy justifies the effort. 

We have computed test patterns to compare forms a) and c), 
considering NPs of diameter $\approx 10 a$, being this the lower size 
limit of validity of the shape-based approach. 

\begin{figure}[hb]
\caption{(Color online) 
A model spherical cluster Au${}_{3925}$ of 5.0~nm diameter with fcc 
structure 
($a$=0.40786~nm) has been constructed according to 
principle c) of \sref{Sec2}. In this case, as the monoatomic fcc 
Wigner-Seitz unit cell contains one atom, principle c) coincides with 
b).\\ 
a: the powder diffraction pattern: red, exact intensity 
$I^{\mathrm{Debye}}_{p}(q)$
calculated by the Debye function 
\cite{GuinierDEB}; blue dotted, $I^{\mathrm{a}}_{p}(q)$ calculated by 
approach a), 
\eref{eq:intAp3}; green dashed, $I^{\mathrm{c}}_{p}(q)$calculation by 
approach 
c), \eref{eq:intCp3}. All intensities have been calculated on an 
absolute scale and then scaled by the same factor.\\
b: lower line, red, difference 
$I^{\mathrm{Debye}}_{p}(q)-I^{\mathrm{c}}_{p}(q)$; 
middle line, green, difference 
$I^{\mathrm{Debye}}_{p}(q)-I^{\mathrm{a}}_{p}(q)$; 
upper line, blue, the exact powder pattern $\times 1/100$ (Debye 
method) 
for comparison.\\
c: lower line, red, difference 
$I^{\mathrm{Debye}}_{p}(q)-s\,I^{\mathrm{c}}_{p}(q)$ after refining a 
scale factor $s=1.024$; 
middle line, green, difference 
$I^{\mathrm{Debye}}_{p}(q)-sI^{\mathrm{a}}_{p}(q)$after refining a 
scale factor $s=1.027$; 
upper line, blue, the exact powder pattern $\times 1/100$ (Debye 
method) 
for comparison. Note that the c)-type pattern difference is flattened 
while the a)-type retains sharp contributions.\\
d: detail around the (111) peak of the $I^{\mathrm{Debye}}_{p}$ and 
$I^{\mathrm{a}}_{p}$ 
patterns (same coding as in part a) 
after scaling, showing 
a significant peak shift for the $I^{\mathrm{a}}_{p}$ pattern.  
}
{\includegraphics[width=85mm]{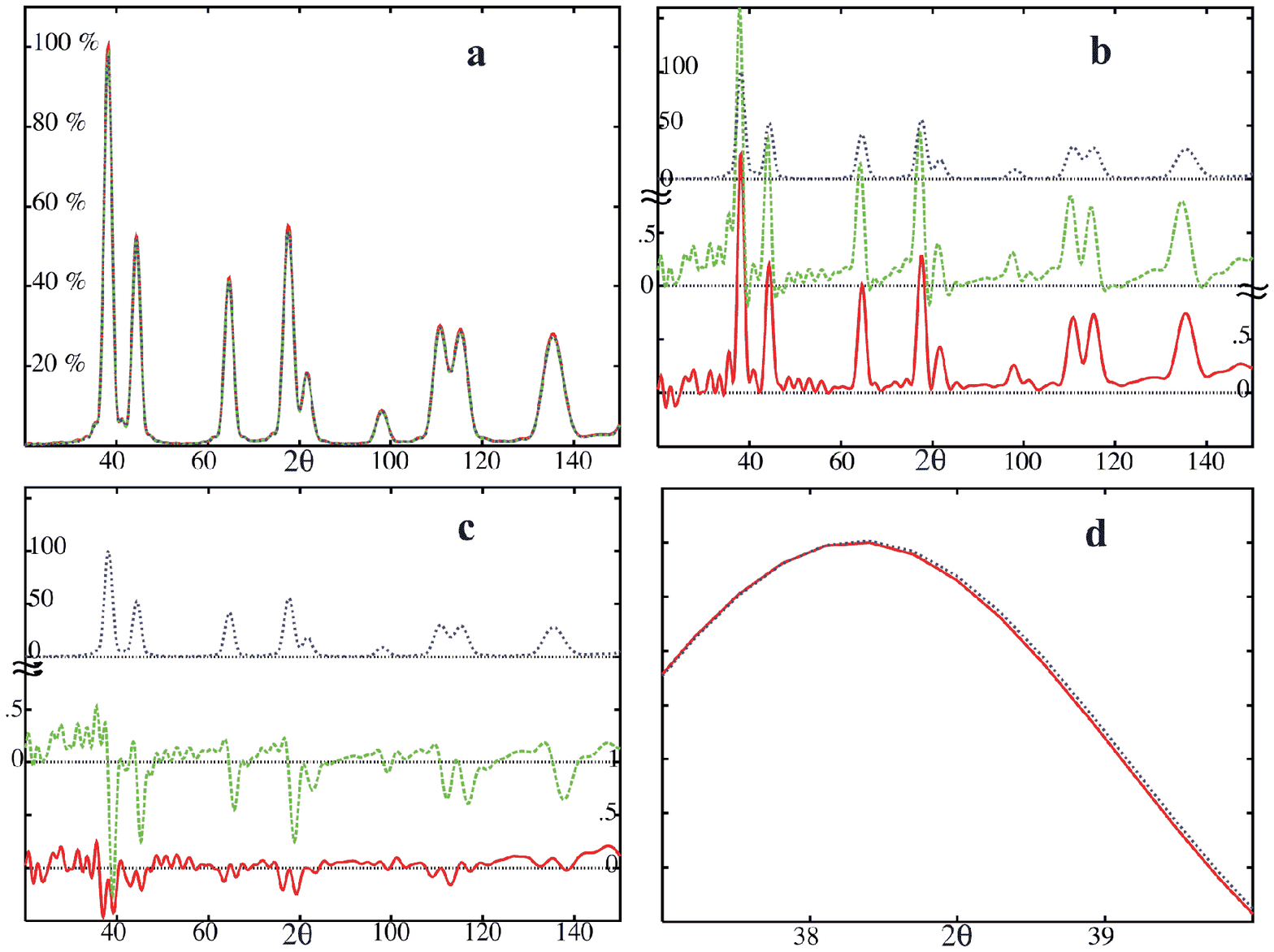}}
%
\label{Fig1}
\end{figure}

We have considered Au spherical NPs of diameter 5~nm ($a$=0.40786~nm, 
$\lambda$=0.154056~nm, 
$2\theta=20\degC\ldots{}150\degC$, 
Lorentz correction and Debye-Waller factor $\exp(-Bq^2/2)$, with 
$B=0.005$~nm${}^2$). 
The powder pattern was calculated exactly by the Debye sum 
\cite{GuinierDEB,NOI03} and 
by \eeref{eq:intAp3}{eq:intCp3}. 
The profiles showed in \fref{Fig1}a are calculated on an absolute 
scale. They 
match quite well, but a maximum error $\approx 2-3\%$ is present in 
both cases a,c. 
The profile $wR$ agreement index between $I^{\mathrm{Debye}}_{p}$ and 
$I^{\mathrm{c}}_{p}$ is 
3.1\%, between $I^{\mathrm{Debye}}_{p}$ and $I^{\mathrm{a}}_{p}$ is 
$wR$=4.4\%. 
The difference profiles (\fref{Fig1}b) show that 
$I^{\mathrm{Debye}}_{p}-I^{\mathrm{c}}_{p}$ has a similar shape to 
$I^{\mathrm{Debye}}_{p}$, while 
$I^{\mathrm{Debye}}_{p}-I^{\mathrm{a}}_{p}$ is quite different. 
Accordingly, 
refining a scale factor between $I^{\mathrm{Debye}}_{p}$ and 
$I^{\mathrm{c}}_{p}$ lowers $wR$ 
to 2.0\% (with featureless difference, \fref{Fig1}c), while a scale 
factor between $I^{\mathrm{Debye}}_{p}$ and
$I^{\mathrm{a}}_{p}$ yields $wR$=3.5\%, with 
still a characteristic difference profile. Furthermore, the peak 
positions 
result very little shifted ($<0.002\degC$) between 
$I^{\mathrm{Debye}}_{p}$ and 
$I^{\mathrm{c}}_{p}$, while they are shifted up to $0.04\degC$ 
between $I^{\mathrm{Debye}}_{p}$ and 
$I^{\mathrm{a}}_{p}$ (\fref{Fig1}d). 

\begin{figure}
\caption{(Color online) 
A model spherical cluster (ZnSe)${}_{1289}$ of 4.8~nm diameter with 
fcc structure 
($a$=0.5633~nm) has been constructed according to 
principle c) of \sref{Sec2}. In this case, as the fcc 
Wigner-Seitz unit cell contains two atoms, construction c) differs 
from  
b).\\ 
a: the powder diffraction pattern: red, exact intensity 
$I^{\mathrm{Debye}}_{p}(q)$
calculated by the Debye function 
\cite{GuinierDEB}; blue dotted, $I^{\mathrm{a}}_{p}(q)$ calculated by 
approach a), 
\eref{eq:intAp3}; green dashed, $I^{\mathrm{c}}_{p}(q)$calculation by 
approach 
c), \eref{eq:intCp3}. Again all intensities have been calculated on 
an 
absoulte scale.\\
b: lower line, red, difference 
$I^{\mathrm{Debye}}_{p}(q)-I^{\mathrm{c}}_{p}(q)$; 
middle line, green, difference 
$I^{\mathrm{Debye}}_{p}(q)-I^{\mathrm{a}}_{p}(q)$; 
upper line, blue, the exact powder pattern $\times 1/100$ (Debye 
method) 
for comparison.\\
c: lower line, red, difference 
$I^{\mathrm{Debye}}_{p}(q)-s\,I^{\mathrm{c}}_{p}(q)$ after refining a 
scale factor $s=0.999$ and an overall-isotropic Debye-Waller factor 
$B=0.468$ ($I^{\mathrm{Debye}}_{p}$ has been evaluated with $B=0.5$); 
middle line, green, difference 
$I^{\mathrm{Debye}}_{p}(q)-sI^{\mathrm{a}}_{p}(q)$ after refining a 
scale factor $s=0.996$ and $B=0.467$; 
upper line, blue, the exact powder pattern $\times 1/100$ (Debye 
method) 
for comparison. Note again that the c)-type pattern difference is 
flattened 
while the a)-type retains sharp contributions. \\
d: detail around the (531) peak of the $I^{\mathrm{Debye}}_{p}$ and 
$I^{\mathrm{a}}_{p}$ 
patterns (same coding as in part a) 
after scaling, showing 
a significant peak shift for the $I^{\mathrm{a}}_{p}$ pattern.  
}
{
{\includegraphics[width=85mm]{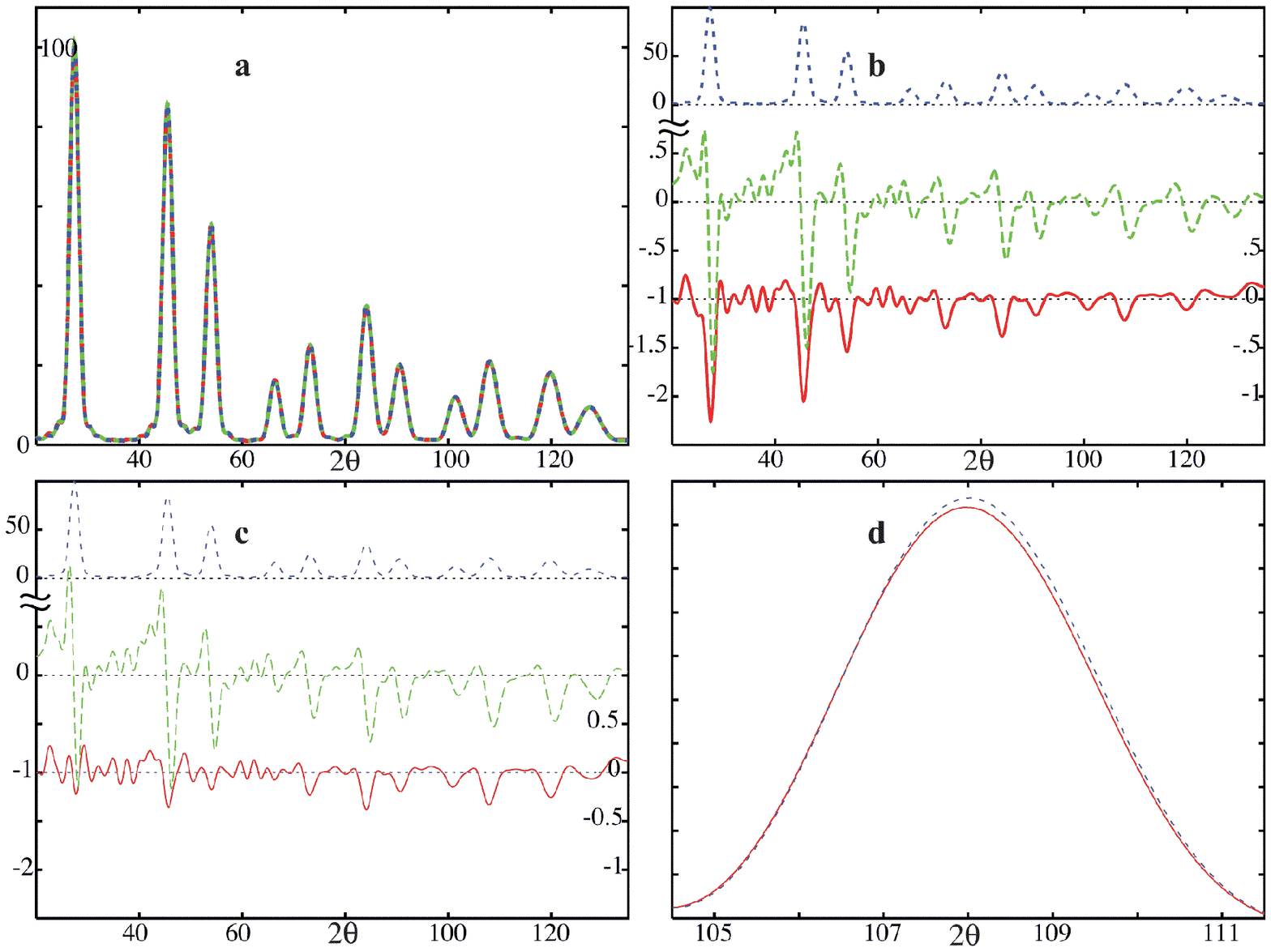}}
}
\label{Fig2}
\end{figure}

Then, 
we have considered ZnSe spherical NPs of diameter 4.8~nm 
($a$=0.5633~nm, 
$\lambda$=0.154056~nm, 
$2\theta=20\degC\ldots{}135\degC$, 
Lorentz correction and Debye-Waller factor with $B=0.005$~nm${}^2$). 
Once more, the powder pattern was calculated exactly by the Debye sum 
\cite{GuinierDEB,NOI03} and 
by \eeref{eq:intAp3}{eq:intCp3}. 
The profiles - calculated on an absoulte scale (\fref{Fig2}a) - 
match quite well with a maximum error 
$\approx 1-2\%$ for both cases a,c. 
The profile agreement index $wR$ between $I^{\mathrm{Debye}}_{p}$ and 
$I^{\mathrm{c}}_{p}$ is 
1.8\%, between $I^{\mathrm{Debye}}_{p}$ and $I^{\mathrm{a}}_{p}$ is 
$wR$=3.1\%. 
The difference profiles (\fref{Fig2}b) show again that 
$I^{\mathrm{Debye}}_{p}-I^{\mathrm{c}}_{p}$ has a similar shape to 
$I^{\mathrm{Debye}}_{p}$, while 
$I^{\mathrm{Debye}}_{p}-I^{\mathrm{a}}_{p}$ is quite different. 
Accordingly, we have  
refined again a scale factor (and this time also a different 
Debye-Waller factor $B$) 
between $I^{\mathrm{Debye}}_{p}$ and 
$I^{\mathrm{c}}_{p}$. $wR$ decreases to 
1.6\% with featureless difference (\fref{Fig2}c). 
On the opposite, when refining 
scale factor and Debye-Waller factor between $I^{\mathrm{Debye}}_{p}$ 
and
$I^{\mathrm{a}}_{p}$ the agreement index does not go below 
$wR$=3.1\%.  
Also the difference profile is little changed (\fref{Fig2}c). Again, 
the peak positions result very little shifted ($<0.001\degC$) between 
$I^{\mathrm{Debye}}_{p}$ and 
$I^{\mathrm{c}}_{p}$, while peak shifts up to $0.05\degC$ between 
$I^{\mathrm{Debye}}_{p}$ and 
$I^{\mathrm{a}}_{p}$ are visible (\fref{Fig2}d). 
Form c) again turns out to be less affected than a) by neglecting the 
cross-term $\mathcal{R}$. 
A small variation of the Debye-Waller factor (from 0.005  to 
0.0047~nm${}^2$) 
is due to the fact that the $\mathcal{R}$-neglection error changes 
slightly the 
intensity ratios. This is however less troublesome than the peak 
shifts observed for form a). 

It results that at NP diameters $D\approx 10 a$ the 
errors in the 
shape-based diffraction pattern calculations, whatever form we 
choose, start to be evident. This approach should not be used below 
this threshold. Also, form a) - which is the standard choice for 
large particles - shows a much larger error and should be avoided in 
favor of c).

\subsection{The log-normal size distribution}\label{SecLN}

There are several 
experimental and theoretical reasons\cite{Kiss99} to believe 
that NP powders have a log-normal distribution of NP size. 
The log-normal distribution of NP radii is usually written in terms 
of its 
mode $R_m$ and width $w_{R}$, as
\begin{equation}
    L(R)=\frac{1}{w_{R}\sqrt{2\pi}}\exp\left\{
    -\frac{[\log(R)-\log(R_m)]^2}{2w_{R}^2}\right\}. 
    \label{eq:lon0}
\end{equation}
The most direct information on a distribution is provided by 
the distribution-averaged NP radius $R_{ave}$ and the relevant 
standard deviation 
$\sigma_{R}$. For a log-normal, the latter parameters are related to 
the former by
\begin{equation}
    {R}_{ave}  =  R_m\exp(w_{R}^2/2);\qquad 
\sigma_{R}=R_m^2\exp(2w_{R}^2);
    \label{eq:rws1}  
\end{equation}
and
\begin{eqnarray}
    {R}_{m} & = & 
\frac{1}{\sqrt{1+\sigma_{R}^2/R_{ave}^2}};\nonumber\\ 
    w_{R}&=&\sqrt{\log\left(1+\sigma_{R}^2/R_{ave}^2\right)}.
    \label{eq:rws2}
\end{eqnarray}
We shall use a form depending directly 
on $R_{ave}$, $\sigma_{R}$ \cite{BalzarRR04}. Setting 
two adimensional parameters $\rho=R/R_{ave}$, 
$c=1+\sigma_{R}^2/R_{ave}^2$, we have
\begin{equation}
    L(R)=\frac{1}{R\sqrt{2\pi\log\left(c\right)}}
    \exp\left[-\frac{\log^2\left(\rho\sqrt{c}\right)}
{2\log\left(c\right)}\right]. 
    \label{eq:lonx}
\end{equation}
Volume- and area-averaged NP diameters can be derived by
\begin{equation}
    D_{V}=\frac{3}{2}R_{ave}c^3;\qquad 
    D_{A}=\frac{4}{3}R_{ave}c^2.
    \label{eq:dadv}
\end{equation}

\section{Nanocrystalline ceria}\label{Sec3}

\subsection{Experimental}\label{sec:EXPE}

X-ray powder diffraction patterns of a nanocrystalline 20-nm 
CeO$_2$ sample, available for a round-robin\cite{BalzarRR04}, 
were downloaded (\verb+http://www.boulder.nist.gov/div853/balzar/+, 
\verb+http://www.du.edu/~balzar/s-s_rr.htm+). The NP size is well 
inside the limits of validity of the shape-based method. 
Among the available datasets, the selected raw data were collected at 
the 
NSLS X3B1 beamline of the Brookhaven National Laboratory in 
flat-plate geometry, 
with a double-crystal Si(111) monochromator on the incident beam 
($\lambda=0.6998$~\AA, $2\theta$ = 12\degC(0.01\degC)60\degC) and a 
Ge(111) analyzer crystal on the diffracted beam.

\subsection{Data preprocessing}

Three data preprocessing stages have been accomplished. 
First, the instrumental function has been deconvoluted by an original 
advanced technique, 
including denoising and background subtraction, 
described in Ref.~\onlinecite{NOI05b}. 

Secondly, the pattern has been fitted by generic asymmetric Voigt 
profiles so 
as to obtain information about peak positions and intensities. 
By comparing the intensities as evaluated from the fit with the 
theoretical ones a small correction for texture and/or 
microabsorption has been evaluated. 
The intensity corrections so obtained have then been stored 
and used in the subsequent stages.

\begin{figure}
\caption{
(Color online) The measured (red error bars) 
and calculated (blue diamonds) lattice 
spacing variations for all well-isolated Bragg peaks of CeO${}_2$ 
plotted against the relevant peaks diffraction angle. 
Error bars have been evaluated assuming a constant error of 
0.0006\degC{} on the anisotropic angular peak shift. 
Calculated values refer to the model of Ref.~\onlinecite{Popa00} 
where 
residual stress components have been refined. 
}
{
{\includegraphics[width=85mm]{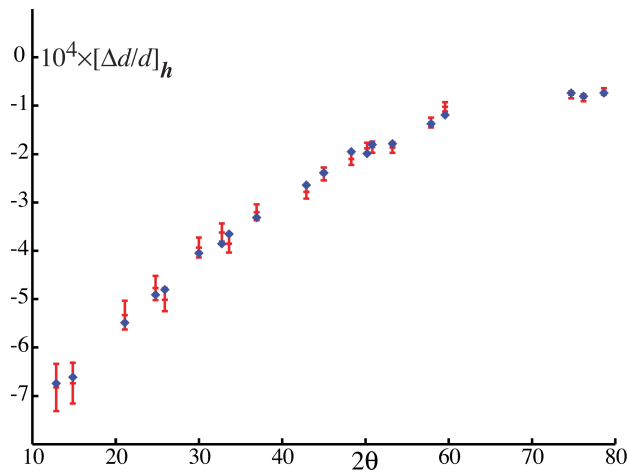}}
}
\label{Fig3}
\end{figure}

Finally, the peak positions were found to be slightly anisotropically 
shifted. 
This has been attributed to a small residual stress, due \emph{e.g.} 
to dislocations. To confirm this point, 
we have evaluated the average lattice spacing variations 
$[\Delta 
d/d]_{\Hv}=-\frac{\pi}{360}\cot(\theta_{\Hv})\Delta(2\theta_{\Hv})$ 
for all single reflections $\Hv$. 
Then we have compared those values with a simple model of elastic 
anisotropy \cite{Popa00}. They resulted in good agreement. In 
\fref{Fig3} we show the 
fit of $[\Delta d/d]_{\Hv}$'s with Eq.~(28) of 
Ref.~\onlinecite{Popa00}. 
The magnitudes of the residual stress 
tensor components, at least for those which can be determined in this 
way, resulted to be in the range 1--10~MPa. 
The values of $\left|\Delta(2\theta_{\Hv})\right|$ are below 
$0.005\degC$, and $[\Delta d/d]_{\Hv}$ range in 1--7$\times 10^{-4}$, 
which are quite small values. As the strain broadening is of the same 
order 
of magnitude  of the peak shifts \cite{Singh01a}, we 
can confirm that strain broadening is rather small in the CeO${}_2$ 
sample 
and can be neglected, as in Ref.~\onlinecite{BalzarRR04}.
Also the residual-stress peak shifts 
so obtained have been saved 
as fixed corrections for the subsequent stages.

\subsection{Full-pattern refinement}\label{sec:REF}

The total intensity diffracted by the powder NP sample is described 
by the sum
\begin{equation}
    I^{cal}(q) = I^{bkg}(q)+\mathop{\sum}_{k=0}^{k_{max}}
    L(R_{k})I_{k}(q),
    \label{eq:ical}
\end{equation}
where $R_{k}=(k+1/2)\Delta R$, $k=1\ldots k_{max}$; 
$I_{k}(q)$ is 
$I^{\mathrm{c}}_{p}(q)$ of \eref{eq:intCp3} evaluated at $R=R_{k}$; 
and $I^{bkg}(q)$ is a polynomial modelling the background. 
The step $\Delta R=a(2\pi/3)^{-1/3}$ is chosen so as to have an 
integer 
number of atoms 
in each $k$-th x-ray sphere of radius $R_{k}$, while keeping the point 
density 
constant and preserving stoichiometry. 
It is evidently possible to use a size-dependent lattice parameter 
$a_{k}$ 
in the calculation of $I_{k}(q)$. For this sample this has been 
deemed unnecessary. Indeed, for diameters of 20~nm, the lattice 
parameter of CeO$_{2}$ has been found\cite{Zhang02} to be already 
equal to the 
bulk value. 
A least-square full-pattern refinement means 
minimizing the quantity 
\begin{equation}
    \chi^2=\mathop{\sum}_{i=1}^{N_{obs}}
    \left(I^{cal}(q_{i})-I^{obs}_i\right)^2w_{i}.
    \label{eq:chi2}
\end{equation}
Here $I^{obs}_i$ is the $i$-th point of the experimental pattern 
corresponding to the scattering vector $q_{i}$, 
$N_{obs}$ the number of experimental points and the weights $w_{i}$
are the estimated inverse variance of the observations. 
The refined parameters are: the
average NPs radius ($R_{ave}$) and the radius dispersion 
$\sigma_{R}$, 
the isotropic Debye-Waller factors $B$ for O and Ce atoms, the
cubic unit cell parameter $a$ and seven background coefficients. 
For the minimization, we have used (for this work) 
a modified simplex algorithm 
\cite{Nelder65}, which is robust but time-consuming; 
however, computing times were reasonable. A derivative-based 
algorithm (Newton, in progress) should give a handsome acceleration. 

\begin{figure}
\caption{
(Color online) Nanosized CeO${}_2$ powder 
pattern final fit. Blue diamonds - the 
observed deconvoluted intensity; red continuous line - the calculated 
intensity; black continuous line, below - difference profile (same 
scale).
}
{
{\includegraphics[width=85mm]{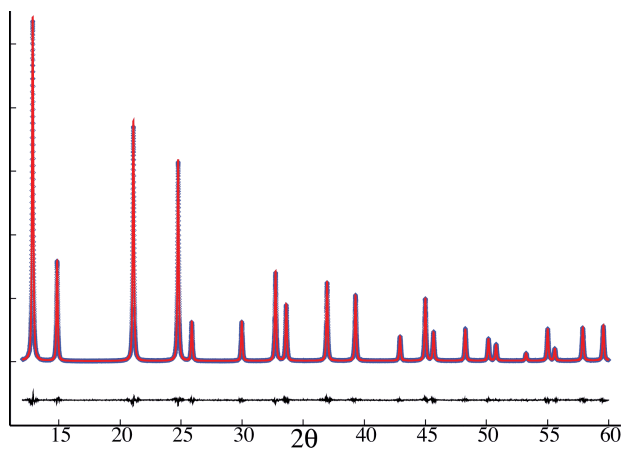}}
}
\label{Fig4}
\end{figure}

The final results are given in \tref{Tab1}, together with the 
corresponding 
values of Ref.~\onlinecite{BalzarRR04}. 
The Debye-Waller factors result to $B_{Ce}=0.0065$~nm$^2$ and 
$B_{O}=0.0084$~nm$^2$. 
The calculated profile is plotted in \fref{Fig4} 
with the experimental pattern and the profile difference. 
The excellent 
fit quality and the final GoF value (1.21) indicate the achievement 
of a reliable result. 
Indeed, the estimated parameters are in good agreement with 
Ref.~\onlinecite{BalzarRR04}. 
The slight discrepancy ($\approx 0.2$~nm), larger than standard 
deviations, might be explained by the improved deconvolution 
method here applied and by the use of the whole pattern instead of a 
limited number 
of peaks as in Ref.~\onlinecite{BalzarRR04}. 

\begin{table}[bth]
    \centering
    \caption{Comparison of size distribution results. 
    Standard deviations are in brackets. Units are nm.}
    \begin{tabular}{l|cccc}
        \hline
        \ &\ \\
&\multicolumn{2}{c}{This 
work}&\multicolumn{2}{c}{Ref.~\onlinecite{BalzarRR04}}\\
        \ &\ \\
        \hline
$R_{ave}$ & 9.58 &(0.02)  & 9.33&(0.07)\\
$\sigma_{R}$ & 4.138 &(0.003)  & 3.92&--\\
$\sigma_{R}^2/R_{ave}^2$ &0.1866 &(0.0008) & 0.177 & 0.003\\
$D_{V}$&24.01&(0.02)&22.8&(0.4)\\
$D_{A}$&17.98&(0.02)&17.2&(0.2)
    \end{tabular}
    \label{Tab1}
\end{table}

\section{Conclusions}\label{Sec4}

The method of shape-convolution to calculate the diffraction 
pattern of NP powders 
has been thoroughly discussed with respect to its 
limits of validity. 
Concerns in applying this method below its optimal size range 
have been demonstrated theoretically and by simulated patterns. 
Finally, 
the effectiveness of full-pattern 
powder data analysis based on the shape-convolution method was proved
to obtain precise size distribution 
information on NP powder samples with a log-normal distribution of 
spherical 
crystallites.


\clearpage\newpage
\setlength{\textwidth}{18cm}
\setlength{\oddsidemargin}{-13mm}

\section*{Figures}

\clearpage\newpage

\clearpage\newpage
\appendix*

\section{Error evaluation}
\label{AppA}
 
Assume to deal with particles of centrosymmetric shape 
$S(\Rv)=S(-\Rv)$ and 
equivalent spherical radius  $R$ (\emph{i.e.}, the radius of the 
sphere of equal volume.). 
The shape Fourier transform $\tilde{S}(\Qv)$ is then a real even 
function:
\begin{equation}
    \tilde{S}(\Qv)=\tilde{S}(-\Qv).
    \label{eq:even}
\end{equation}
Recall also that the  
gradient of an even function is odd:
\begin{equation}
    \tilde{\VK{G}}(\Qv)\equiv\nabla_{\Qv}\tilde{S}(\Qv)=-\tilde{\VK{G}}(-\Qv).
    \label{eq:oddG}
\end{equation}

Our aim is to evaluate - for the different forms a), b), c) as 
introduced 
in \sref{Sec2} and carried out in \sref{Sec22}, \sref{Sec23} - 
the neglected residual intensity contribution 
$\mathcal{R}(\Qv)$ of \eref{eq:X} with respect 
to the respective retained term (\emph{cf.} 
\eeeref{eq:intA}{eq:intB}{eq:intC}) in the immediate 
vicinity of a Bragg peak. 

Let  $\Hv_{0}$ the nearest Bragg peak to $\Qv$. 
First note that, if $|\Qv-\Hv_{0}| \gg 1/R$, 
$\mathcal{R}(\Qv)$ is of order $(2\pi qR)^{-4}$, so we neglect it 
altogether. 
If $\Qv$ is very close to $\Hv_{0}$, set $\Qv=\Hv_{0}+ 
\Delta\Qv$ (so $\Delta q \lesssim 1/R$). 
We can drop in the sum over $\Hv$ all terms with $\Hv\neq\Hv_{0}$ 
because 
they are $O(2\pi qR)^{-4}$ and 
reorder the second sum, 
obtaining 
\begin{eqnarray}
     \mathcal{R}^{\mathrm{a}}(\Hv_{0}+\Delta\Qv)
    &\approx&
    \tilde{S}(\Delta\Qv)
     \mathop{\sum}_{\genfrac{}{}{0pt}{2}{\Kv\in\Lambda^*}{\Kv\neq 0\ 
}}
     \tilde{S}(\Delta\Qv+\Kv) F(\Hv_{0})\times\nonumber\\
&\times&\overline{F}(\Hv_{0}-\Kv)
    \label{eq:AA1}  \end{eqnarray}
\begin{eqnarray}
     \mathcal{R}^{\mathrm{b}}(\Hv_{0}+\Delta\Qv)
    &\approx&
    \tilde{S}(\Delta\Qv)
    \left|F(\Hv_{0}+\Delta\Qv)\right|^2\times\nonumber\\&\times&
     \mathop{\sum}_{\genfrac{}{}{0pt}{2}{\Kv\in\Lambda^*}{\Kv\neq 0\ 
}}
     \tilde{S}(\Delta\Qv+\Kv) 
    \label{eq:AA2}  \end{eqnarray}
\begin{eqnarray}
    \mathcal{R}^{\mathrm{c}}(\Hv_{0}+\Delta\Qv)
    &\approx&
    \tilde{S}(\Delta\Qv)
     \mathop{\sum}_{\genfrac{}{}{0pt}{2}{\Kv\in\Lambda^*}{\Kv\neq 0\ 
}}
     \tilde{S}(\Delta\Qv+\Kv) \times\nonumber\\&
\times&F(\Hv_{0},q)\overline{F}(\Hv_{0}-\Kv,q)
    \label{eq:AA3}  \\
    \text{where}&&q\equiv|\Hv_{0}+\Delta\Qv|\nonumber
\end{eqnarray}
At the same time, for 
$\Qv=\Hv_{0}+\Delta\Qv$ with $\Delta q \lesssim 1/R$, the 
intensities $I^{\mathrm{x}}(\Qv)$ 
of \eeeref{eq:intA}{eq:intB}{eq:intC} can be approximated 
by the $\Hv_{0}$-th term of the RHS sum, neglecting 
terms of $O(2\pi qR)^{-4}$. 
Furthermore, in general, $\tilde{S}(\Delta\Qv)=\tilde{S}(0)+O(\Delta q^2)$. 
Therefore, the ratios 
$\mathcal{R}^{\mathrm{x}}(\Hv_{0}+\Delta\Qv)/I^x(\Hv_{0}+\Delta\Qv)$ 
are given by 
\begin{eqnarray}
    \hat{\mathcal{R}}^{\mathrm{a}}(\Hv_{0},\Delta\Qv)
    &\equiv&\frac{\mathcal{R}^{\mathrm{a}}(\Hv_{0}+\Delta\Qv)}
    {\tilde{S}^2(\Delta\Qv)
    |F(\Hv_{0})|^2}\nonumber\\
    &\approx&
     \mathop{\sum}_{\genfrac{}{}{0pt}{2}{\Kv\in\Lambda^*}{\Kv\neq 0\ }}
     \tilde{s}(\Delta\Qv,\Kv) \frac{\overline{F}(\Hv_{0}-\Kv)}
     {\overline{F}(\Hv_{0})}
    \label{eq:AAr1}  \end{eqnarray}
\begin{eqnarray}
    \hat{\mathcal{R}}^{\mathrm{b}}(\Hv_{0},\Delta\Qv)
    &\equiv&\frac{\mathcal{R}^{\mathrm{b}}(\Hv_{0}+\Delta\Qv)}
    {\tilde{S}^2(\Delta\Qv)
    |F(\Hv_{0}+\Delta\Qv)|^2}\nonumber\\
    &\approx&
     \mathop{\sum}_{\genfrac{}{}{0pt}{2}{\Kv\in\Lambda^*}{\Kv\neq 0\ }}
     \tilde{s}(\Delta\Qv,\Kv) 
    \label{eq:AAr2}  \end{eqnarray}
\begin{eqnarray}
    \hat{\mathcal{R}}^{\mathrm{c}}(\Hv_{0},\Delta\Qv)
    &\equiv&\frac{\mathcal{R}^{\mathrm{c}}(\Hv_{0}+\Delta\Qv)}
    {\tilde{S}^2(\Delta\Qv)
    |F(\Hv_0,q)|^2}\nonumber\\
    &\approx&
     \mathop{\sum}_{\genfrac{}{}{0pt}{2}{\Kv\in\Lambda^*}{\Kv\neq 0\ }}
     \tilde{s}(\Delta\Qv,\Kv)\frac{\overline{F}(\Hv_{0}-\Kv,q)}
     {\overline{F}(\Hv_{0},q)}
    \label{eq:AAr3}  \\
    \text{where}&&\tilde{s}(\Delta\Qv,\Kv)
    \equiv\frac{\tilde{S}(\Delta\Qv+\Kv)
    }{\tilde{S}(0)}\nonumber
\end{eqnarray}
Note that, because of \eeref{eq:even}{eq:oddG}, we have
\begin{eqnarray}
    \tilde{s}(\Delta\Qv,\Kv)&=&\tilde{s}(-\Delta\Qv,-\Kv);
    \label{eq:even2}\\
    \tilde{\VK{g}}(\Delta\Qv,\Kv)&\equiv&\nabla_{\Delta\Qv}
    \tilde{s}(\Delta\Qv,\Kv)=\tilde{\VK{G}}
(\Delta\Qv+\Kv)/\tilde{S}(0)\nonumber\\
    &=&-\tilde{\VK{g}}(-\Delta\Qv,-\Kv).
    \label{eq:oddg}
\end{eqnarray}
We can immediately veryfy that in case b) it results
\begin{equation}
    \nabla_{\Delta\Qv}\hat{\mathcal{R}}^{\mathrm{b}}(\Hv_{0},\Delta\Qv)=
     \mathop{\sum}_{\genfrac{}{}{0pt}{2}{\Kv\in\Lambda^*}{\Kv\neq 0\ 
}}
     \tilde{\VK{g}}(\Delta\Qv,\Kv)
    \label{eq:prB1}
\end{equation}
In the sum above 
the term with index $\Kv$ is always accompanied by a term with 
index $-\Kv$. Setting also $\Delta\Qv=0$, and using \eref{eq:oddg}, 
we have
\begin{equation}
    \nabla_{\Delta\Qv}\hat{\mathcal{R}}^{\mathrm{b}}(\Hv_{0},0)=
     \mathop{\sum}_{\Kv\in\Lambda^*/2}
     \left(\tilde{\VK{g}}(0,\Kv)+\tilde{\VK{g}}(0,-\Kv)\right)=0\,.
    \label{eq:prB2}
\end{equation}
where $\Lambda^*/2$ denotes an arbitrarily chosen half-space of the 
reciprocal lattice 
without the origin. 
Now, expanding 
$\hat{\mathcal{R}}^{\mathrm{b}}(\Hv_{0},\Delta\Qv)$ 
in Taylor series at $\Delta\Qv=0$, 
we have
\begin{equation}
    \hat{\mathcal{R}}^{\mathrm{b}}(\Hv_{0},\Delta\Qv)
    \approx\hat{\mathcal{R}}^{\mathrm{b}}(\Hv_{0},0)+O(\Delta q^2).
    \label{eq:taylB}
\end{equation}
Note also in \eref{eq:AAr2} that 
$\hat{\mathcal{R}}^{\mathrm{b}}(\Hv_{0},0)$ does 
not depend on the considered Bragg reflection $\Hv_0$. Therefore, 
we can write
\begin{equation}
    {\mathcal{R}}^{\mathrm{b}}(\Qv)\propto I^{\mathrm{b}}(\Qv),
    \label{eq:finalb}
\end{equation}
and the proportionality constant can be evaluated by \eref{eq:AAr2} 
with $\Delta\Qv=0$. 
We can conclude that 
the effect of neglecting ${\mathcal{R}}^{\mathrm{b}}$ will 
be just a relative error on 
the global profile scale factor. This factor is size-dependent, 
however, therefore for size distribution analysis at 
small sizes it may be necessary to 
introduce a correction as from \eref{eq:AAr2}.

Cases a), c), are more complex. 
We are interested to powder diffraction, where 
    $I(\Qv)$ is to be integrated at constant $q$, 
    therefore we shall consider
\begin{equation}
    \overline{\mathcal{R}}^{\mathrm{x}}(\Hv_{0},\Delta\Qv)
    =\frac{1}{2}\left[\hat{\mathcal{R}}^{\mathrm{x}}(\Hv_{0},\Delta\Qv)
    +\hat{\mathcal{R}}^{\mathrm{x}}(-\Hv_{0},-\Delta\Qv)\right]
    \label{eq:topo}
\end{equation}
Expanding $\overline{\mathcal{R}}^{\mathrm{x}}(\Hv_{0},\Delta\Qv)$ in 
Taylor 
series at $\Delta\Qv=0$, 
we have
\begin{equation}
    \overline{\mathcal{R}}^{\mathrm{x}}(\Hv_{0},\Delta\Qv)
    \approx\overline{\mathcal{R}}^{\mathrm{x}}(\Hv_{0},0)
    +\nabla_{\Delta\Qv}\overline{\mathcal{R}}^{\mathrm{x}}(\Hv_{0},0)
    \cdot\Delta\Qv+O(\Delta q^2).
    \label{eq:tayl}
\end{equation}
We shall now develop 
$\overline{\mathcal{R}}^{\mathrm{a}}(\Hv_{0},\Delta\Qv)$ and 
$\nabla_{\Delta\Qv}\overline{\mathcal{R}}^{\mathrm{a}}(\Hv_{0},0)$ 
in cases a, c.

\subsection{Case c)}

First, recall that the atomic form factors $f_{\alpha}(q)$ are 
constants for neutron scatering and monotonically decreasing smooth 
functions in the x-ray case. 
In the latter case, furthermore, the form factors of different 
elements 
have remarkably 
similar profiles. For a structure with $N_{a}$ atoms in the unit 
cell, 
it is then possible \cite{GiacU} to approximate
\begin{equation}
    f_{\alpha}(q)\approx c_{\alpha}\langle f(q)\rangle\equiv 
    c_{\alpha}\frac{1}{N_{a}}\mathop{\sum}_{\beta=1}^{N_a}f_{\beta}(q).
    \label{eq:aff}
\end{equation}
with $c_{\alpha}$ appropriate constants. 
Therefore the structure factor ratios appearing in \eref{eq:AAr3} can 
be 
simplified as
\begin{equation}
    \frac{\overline{F}(\Hv_{0}-\Kv,q)}{\overline{F}(\Hv_{0},q)}\approx
    \frac{\mathop{\sum}_{\alpha=1}^{N_{a}}c_{\alpha}{\mathrm{e}}^{-2\pi 
    i(\Hv_{0}-\Kv)\cdot\Rv_{\alpha}}}
    {\mathop{\sum}_{\beta=1}^{N_{a}}c_{\beta}{\mathrm{e}}^{-2\pi 
    i\Hv_{0}\cdot\Rv_{\beta}}}\equiv \tau(\Hv_{0},\Kv),
    \label{eq:fapp}
\end{equation}
independent of $q=|\Hv_{0}+\Delta\Qv|$. 
Note that
\begin{equation}
    \tau(-\Hv_{0},-\Kv)=\overline{\tau}(\Hv_{0},\Kv); \qquad
    \tau(-\Hv_{0},\Kv)=\overline{\tau}(\Hv_{0},-\Kv).
    \label{eq:taup}
\end{equation}
Now we can write  explicitly $\overline{\mathcal{R}}^{\mathrm{c}}$ 
using \eeref{eq:AAr3}{eq:topo} and 
\begin{eqnarray}
    \overline{\mathcal{R}}^{\mathrm{c}}(\Hv_{0},\Delta\Qv)
    &=&\frac{1}{2}
    \mathop{\sum}_{\genfrac{}{}{0pt}{2}{\Kv\in\Lambda^*}{\Kv\neq 0\ }}
     \left[\tilde{s}(\Delta\Qv,\Kv)
\tau(\Hv_{0},\Kv)+\right.\nonumber\\&&\left.+\tilde{s}(-\Delta\Qv,\Kv)
\overline{\tau}(\Hv_{0},-\Kv)\right].
    \label{eq:xpl1}
\end{eqnarray}
Splitting the sum, reordering $\Kv\rightarrow -\Kv$ in one part, 
using \eref{eq:even2} 
and recombining, 
we have
\begin{equation}
    \overline{\mathcal{R}}^{\mathrm{c}}(\Hv_{0},\Delta\Qv)
    =
    \mathop{\sum}_{\genfrac{}{}{0pt}{2}{\Kv\in\Lambda^*}{\Kv\neq 0\ }}
     \tilde{s}(\Delta\Qv,\Kv){\mathrm{Re}}\left[
     \tau(\Hv_{0},\Kv)\right].
    \label{eq:xpl}
\end{equation}
Again as in \eref{eq:prB2}, we can pair terms with $\Kv$ and $-\Kv$. 
Using \eref{eq:taup}, we obtain
\begin{eqnarray}
    \overline{\mathcal{R}}^{\mathrm{c}}(\Hv_{0},\Delta\Qv)
    &=&
    \mathop{\sum}_{{\Kv\in\Lambda^*/2}}
     {\mathrm{Re}}\left[\tilde{s}(\Delta\Qv,\Kv)
\tau(\Hv_{0},\Kv)+\right.\nonumber\\&&\left.+\tilde{s}
(\Delta\Qv,-\Kv)\tau(\Hv_{0},-\Kv)\right].
    \label{eq:xpl2}
\end{eqnarray}
Define now the arbitrary 
half-lattice $\Lambda^*/2$ as that defined by a plane $\perp\Hv_{0}$ 
passing through the origin 
and containing $\Hv_{0}$. The origin is excluded. 
We have 
\begin{equation}
    \overline{\mathcal{R}}^{\mathrm{c}}(\Hv_{0},0)
    =
    \mathop{\sum}_{{\Kv\in\Lambda^*/2}}\tilde{s}(0,\Kv)
     {\mathrm{Re}}\left[
\tau(\Hv_{0},\Kv)+\tau(\Hv_{0},-\Kv)\right].
    \label{eq:xplf}
\end{equation}
Then, evaluating the gradient in $\Delta\Qv=0$, using 
\eref{eq:oddG}, we have finally
\begin{eqnarray}
    \nabla_{\Delta\Qv}\overline{\mathcal{R}}^{\mathrm{c}}(\Hv_{0},0)
    &=&
    \mathop{\sum}_{{\Kv\in\Lambda^*/2}}
     \tilde{\VK{g}}(0,\Kv)\times\nonumber\\&\times&
     {\mathrm{Re}}\left[\tau(\Hv_{0},\Kv)-\tau(\Hv_{0},-\Kv)\right].
    \label{eq:xplg}
\end{eqnarray}
The gradient $\nabla_{\Delta\Qv}\overline{\mathcal{R}}^{\mathrm{c}}(\Hv_{0},0)$ 
is a vector. 
We have to take its angular average to 
determine the effect on the powder pattern. 
This is done by simply 
taking the 
scalar product with $\Hvh_0\equiv\Hv_0/h_0$:
\begin{eqnarray}
    \nabla_{\Delta\Qv}\overline{\mathcal{R}}^{\mathrm{c}}(\Hv_{0},0)
    \cdot\Hvh_0
    &=&
   \mathop{\sum}_{{\Kv\in\Lambda^*/2}}
     (\tilde{\VK{g}}(0,\Kv)\cdot\Hvh_0)\times\nonumber\\&\times&
     {\mathrm{Re}}\left[\tau(\Hv_{0},\Kv)-\tau(\Hv_{0},-\Kv)\right].
    \label{eq:palle}
\end{eqnarray}

For spherical shape, it will be $\tilde{\VK{g}}(0,\Kv)||\Kv$;
therefore terms with $\Kv\perp\Hv_0$ will be zero and 
those with $\Kv||\Hv_0$ will be most important. 
Both $\tilde{\VK{g}}(0,\Kv)$ and $\tilde{s}(0,\Kv)$ are damped 
oscillatory functions with amplitude $\sim (2\pi kR)^{-2}$. 
As $1/a\lesssim k$, the magnitudes of both 
$\overline{\mathcal{R}}^{\mathrm{c}}(\Hv_{0},0)$ and 
$\nabla_{\Delta\Qv}\overline{\mathcal{R}}^{\mathrm{c}}(\Hv_{0},0)$
are of order $(a/R)^{2}$. 
Unfortunately, \eref{eq:palle} cannot be estimated more in detail, 
because of the dependence from the `reduced' structure factors 
$\tau(\Hv_{0},\Kv)$. 
However, we can assess that its importance would be smaller than 
the corresponding term for case a) for x-ray scattering.

\subsection{Case a)}

In case a), we can trace the same steps as in case c) 
but instead of the `reduced' structure factors $\tau(\Hv_{0},\Kv)$ we 
have to consider 
the ratios
\begin{equation}
    \zeta(\Hv_{0},\Kv) = 
\frac{\overline{F}(\Hv_{0}-\Kv)}{\overline{F}(\Hv_{0})} =
    \frac{\overline{F}(\Hv_{0}-\Kv)F(\Hv_{0})}
    {\left|{F}(\Hv_{0})\right|^2}
    \label{eq:casea1}
\end{equation}
and in the analog sums of \eref{eq:xplf} and \eref{eq:palle} 
for 
$\overline{\mathcal{R}}^{\mathrm{a}}(\Hv_{0},0)$ and 
$\nabla_{\Delta\Qv}\overline{\mathcal{R}}^{\mathrm{a}}(\Hv_{0},0)$
there will appear terms as 
\begin{equation}
    {\mathrm{Re}}\left[\zeta(\Hv_{0},\Kv)
    \pm\zeta(\Hv_{0},-\Kv)\right]
    .
    \label{eq:casea2}
\end{equation}
The most important terms for the powder pattern are again those 
with $\Kv||\Hv_0$. The structure factors $F(\Hv_{0}\pm\Kv)$ 
(see \eref{eq:STRF}) 
depend on form factors $f_{\alpha}(|\Hv_{0}\pm\Kv|)$, 
and for $\Kv||\Hv_0$ these will be strongly different. 
This in turn will amplify the differences $\xi(\Hv_{0},\Kv)$. 
Therefore it is likely that for case a) the effect of the neglected 
term ${\mathcal{R}}^{\mathrm{a}}$ will be significantly larger than 
for case c). The examples reported in \sref{Sec23} show just that.

\end{document}